\documentclass[12pt]{article}
\usepackage[english]{babel}
\usepackage{epsfig}
\begin{document}
\large

\begin{center}
{\bf A remark concerning the standard approach to $CP$ violation in a
system of $K^o$ mesons} \vspace{0.3cm}

Kh. M. Beshtoev \vspace{0.1cm}
\par
beshtoev@lxmx00.jinr.ru
\par
\vspace{0.3cm} Joint Institute for Nuclear Research, Joliot Curie
6, 141980 Dubna, Moscow region, Russia.
\end{center}

\par
\begin{center}
Abstract
\end{center}

\par
\noindent

Within the standard approach to $CP$ violation in
a system of $K^o$ mesons, the normalization factor in the expression
for the transition probability $|K^o_1|^2$ contains the $CP$
violation phase. A normalization multiplier for the transition
probability can obviously not contain a phase term. In this work
two simple methods are proposed for resolving this issue.


\section{Introduction}

\par
Parity $P$ was previously supposed to be a good number;
however, after theoretical \cite{1} and experimental \cite{2}
works it has become clear that $P$ parity is violated in weak
interactions. Then, in ref. \cite{3} an assumption was put
forward that, although $P$ parity is not conserved in weak
interactions, $CP$ parity is conserved. Ref. \cite{4} reported
that with a probability of about $0.2\%$ there exists in $K_L$
decays a two $\pi$ decay mode which actually points to $CP$
parity violation.
\par
In all textbooks and monographs, where the problem of $CP$ violation
in a system of $K^o$ mesons is dealt with, the primary
$K^o$ and $\bar K^o$ mesons are, first, considered to transform under
the violation of strangeness $S$, into a superposition of $K^o_1$ and
$K^o_2$ mesons and, then, under $CP$ violation these $K^o_1$ and $K^o_2$
mesons transform into a superposition of $K_S, K_L$ mesons \cite{5}
$$
\begin{array}{cc} K_S = \frac{1}{\sqrt{1 + |\varepsilon|^2}}
(K^o_1 + \varepsilon K^o_2 ),\\ K_L = \frac{1}{\sqrt{1 +
|\varepsilon|^2}} (\varepsilon K^o_1 + K^o_2) .\end{array}
  \eqno(1)
$$
where $\varepsilon$ is complex value. Inverse transformation of
expressions (1) results in
$$
\begin{array}{cc} K^o_1 = \frac{\sqrt{1 + |\varepsilon|^2}}{1-
\varepsilon^2} (K_S - \varepsilon K_L ),\\ K^o_2 = \frac{\sqrt{1 +
|\varepsilon|^2}}{1- \varepsilon^2}  (-\varepsilon K_S + K_L)
.\end{array}
  \eqno(2)
$$
We can rewrite $\varepsilon$ as $\varepsilon =
|\varepsilon| e^{- i \delta}$. Then, taking into account that
$K_S(t) = e^{(-iE_S-\Gamma_S/2) t} K_S(0)$, $K_L(t) = e^{(-iE_L
-\Gamma_L/2) t} K_L(0)$, we obtain
$$
|K^o_1(t)|^2 = \frac{{1 + |\varepsilon|^2}}{|1- \varepsilon^2|^2}
[e^{\Gamma_S t} + |\varepsilon|^2 e^{\Gamma_L t} - 2 |\varepsilon|
e^{\frac{\Gamma_S + \Gamma_L}{2} t}cos((E_L-E_S) t)] =
$$
$$
\frac{{1 + |\varepsilon|^2}}{(1+ |\varepsilon|^4 - 2
|\varepsilon|^2 cos \delta)} [e^{\Gamma_S t} + |\varepsilon|^2
e^{\Gamma_L t} - 2 |\varepsilon| e^{\frac{\Gamma_S + \Gamma_L}{2}
t}cos((E_L-E_S) t + \delta)] . \eqno(3)
$$
$$
|K^o_2(t)|^2 = \frac{{1 + |\varepsilon|^2}}{(1+ |\varepsilon|^4 -
2 |\varepsilon|^2 cos \delta)} [|\varepsilon|^2 e^{\Gamma_S t} +
e^{\Gamma_L t} - 2 |\varepsilon| e^{\frac{\Gamma_S + \Gamma_L}{2}
t}cos((E_L-E_S) t - \delta)] . \eqno(3')
$$

We see that the normalization factor in (3) contains the phase
term $cos \delta$. Evidently, this phase term is not related to
the normalization of states. Therefore, we have to get rid of this
term. We shall further deal with such an approach.
\par

\section{$CP$ violation without phase term in normalization factor in
expression for transition probability}

To avoid the presence of a phase term in the normalization factor
for the transition probability we replace $\varepsilon$ in the
second term of (1) by $\varepsilon^{*} = |\varepsilon| e^{i \delta}$:
$$
K_S = \frac{1}{\sqrt{1 + |\varepsilon|^2}} (K^o_1 + \varepsilon
K^o_2 ) \equiv \frac{1}{\sqrt{1 + |\varepsilon|^2}} (K^o_1 +
|\varepsilon| e^{-i \delta} K^o_2 )
$$
$$
K_L = \frac{1}{\sqrt{1 + |\varepsilon^{*}|^2}} (\varepsilon^{*}
K^o_1 + K^o_2) \equiv \frac{1}{\sqrt{1 + |\varepsilon|^2}}
(|\varepsilon| e^{i \delta} K^o_1 + K^o_2) . \eqno(4)
$$
By inverse transformation of expression (4) we obtain
$$
\begin{array}{cc} K^o_1 = \frac{\sqrt{1 + |\varepsilon|^2}}{1-
|\varepsilon|^2} (-K_S + |\varepsilon| e^{-i \delta} K_L ),\\
K^o_2 = \frac{\sqrt{1 + |\varepsilon|^2}}{1- |\varepsilon|^2}
(|\varepsilon| e^{i \delta} K_S - K_L) .\end{array}
  \eqno(5)
$$
Then, for $|K^o_1(t)|^2$ and $|K^o_2(t)|^2$ we get
$$
|K^o_1(t)|^2 = \frac{{1 + |\varepsilon|^2}}{1- |\varepsilon^2|^2}
[e^{\Gamma_S t} + |\varepsilon|^2 e^{\Gamma_L t} - 2 |\varepsilon|
e^{\frac{\Gamma_S + \Gamma_L}{2} t}cos((E_L-E_S) t + \delta)] .
\eqno(6)
$$
$$
|K^o_2(t)|^2 = \frac{{1 + |\varepsilon|^2}}{1- |\varepsilon^2|^2}
[|\varepsilon|^2 e^{\Gamma_S t} + e^{\Gamma_L t} - 2 |\varepsilon|
e^{\frac{\Gamma_S + \Gamma_L}{2} t}cos((E_L-E_S) t + \delta)] .
\eqno(6')
$$
We can go farther and use the following expressions for the $ K_S,
K_L$ states:
$$
K_S = \frac{1}{\sqrt{1 + |\varepsilon|^2}} (K^o_1 + \varepsilon
K^o_2 ) \equiv \frac{1}{\sqrt{1 + |\varepsilon|^2}} (K^o_1 +
|\varepsilon| e^{-i \delta} K^o_2 )
$$
$$
K_L = \frac{1}{\sqrt{1 + |\varepsilon^{*}|^2}} (-\varepsilon^{*}
K^o_1 + K^o_2) \equiv \frac{1}{\sqrt{1 + |\varepsilon|^2}}
(-|\varepsilon| e^{i \delta} K^o_1 + K^o_2) . \eqno(7)
$$
where $- \varepsilon^{*}$ is substituted for $\varepsilon$ in the
second term of expressions (1).
\par
By inverse transformation we obtain
$$
\begin{array}{cc} K^o_1 =
(K_S - |\varepsilon| e^{-i \delta} K_L ),\\ K^o_2 =
 (|\varepsilon| e^{i \delta} K_S +  K_L) .\end{array}
  \eqno(8)
$$
Then, for $|K^o_1(t)|^2$ we get
$$
|K^o_1(t)|^2 = [e^{\Gamma_S t} + |\varepsilon|^2 e^{\Gamma_L t} -
2 |\varepsilon| e^{\frac{\Gamma_S + \Gamma_L}{2} t}cos((E_L-E_S) t
+ \delta)] . \eqno(9)
$$

\section{Conclusion}

Within the standard approach to $CP$ violation in a system of
$K^o$ mesons,  the normalization factor in the expression for the
transition probability $|K^o_1|^2$ contains a $CP$ violation
phase. A normalization multiplier for the transition probability
can obviously not contain a phase term. In the present work, two
simple methods are put forward for resolving this issue. To this
end two approaches are applied. In the first approach
$\varepsilon^{*}$ is substituted for the term $\varepsilon$ in the
expression for $K_L$, (1), while in the second $- \varepsilon^{*}$
is substituted for $\varepsilon$ in expression (1) for $K_L$.
Then, the renormalization factor in the expression for $|K^o_1|^2$
no longer contains any phase term, i.e. no $CP$ violation phase
term is present in the normalization factor of the expression for
the transition probability $|K^o_1|^2$.
\par
So we see that in expressions for transition probabilities
$|K^o_1(t)|^2$ and $|K^o_2(t)|^2$ the phase term $\delta$ has
different signs (see expr. (3) and (3')) while these probabilities
have the same sign in our approach (see expr. (6) and (6')).
Besides difference between old and our normalization factor is
$\Delta N= \frac{{1 + |\varepsilon|^2}}{(1+ |\varepsilon|^4 - 2
|\varepsilon|^2 cos \delta)} - \frac{{1 + |\varepsilon|^2}}{1-
|\varepsilon^2|^2}$ $\simeq 2 |\varepsilon|^2 cos \delta)$ (where
$|\varepsilon|=2.23 \cdot 10^{-3}$ (see ref. \cite{5}) and it is
very small value).
\par
It is necessary to remark that in experiment with high precision
we can fulfill examination of normalization factor and sign of
phase factor $\alpha$ in order to determine which of the above two
approaches is realized indeed.



\begin{thebibliography}{999}
\par
\bibitem{1} T. D. Lee, C. N. Yang, Phys. Rev. 104(1956)254.
\par
\bibitem{2} C. S. Wu et al., Phys. Rev. 105(1957)1413;
\par
Phys. Rev. 106(1957)1361.
\par
\bibitem{3} L. D. Landau, Soviet J.JETP 32(1957)405.
\par
\bibitem{4} J. H. Christenson et al., Phys. Rev. Lett. 13(1964)138.
\par
\bibitem{5} M. A. Tomson, Michaelmas Term 2009, p.441.

\end{thebibliography}
\end{document}